\documentclass[12pt]{article}

\usepackage{times}
\usepackage{fixltx2e}
\usepackage{graphicx}
\usepackage[table]{xcolor}
\usepackage{ragged2e}
\usepackage{multirow}
\usepackage{helvet}
\usepackage{booktabs}
\usepackage{colortbl}
\usepackage{amsmath}
\usepackage{cmbright}

\usepackage{hyperref}
\hypersetup{
    colorlinks=true,
    linkcolor=magenta,
    filecolor=magenta, 
    citecolor=sared,
    urlcolor=red,
}

\usepackage{blindtext}

\makeatletter
\newenvironment{figurehere}
{\def\@captype{figure}}
{}
\makeatletter

\usepackage{ccaption}
\usepackage{siunitx}
\usepackage{fixltx2e}
\newenvironment{sciabstract}{%
\begin{quote} \bf}
{\end{quote}}

\definecolor{sared}{rgb}{0.69, 0,0}

\newcounter{lastnote}

\title{\textbf{A Bio-inspired Asymmetric Double-Gate Ferroelectric FET for Emulating Astrocyte and Dendrite Dynamics in Neuromorphic Systems}} 

\author
{Zhouhang Jiang$^{1 \dag}$, A N M Nafiul Islam$^{2 \dag}$, Zhuangyu Han$^{2}$, Zijian Zhao$^{1}$, \\Franz M{\"u}ller$^{3}$, Jiahui Duan$^{1}$,
Halid Mulaosmanovic$^{4}$, Stefan D{\"u}nkel$^{4}$, \\ Sven Beyer$^{4}$,  Sourav Dutta$^{5}$, Vijaykrishnan Narayanan$^{2}$, Thomas K{\"a}mpfe$^{3,6}$, \\ Suma George Cardwell$^{7}$, Frances Chance$^{7}$, Abhronil Sengupta$^{2*}$, \\ Kai Ni$^{1*}$
\\
\\
\normalsize{$^{1}$University of Notre Dame, Notre Dame 46556, USA}\\
\normalsize{$^{2}$Pennsylvania State University, State College, PA 16802, USA}\\
\normalsize{$^{3}$Fraunhofer IPMS, Dresden, Germany}\\
\normalsize{$^{4}$GlobalFoundries Fab1 LLC \&Co. KG, Dresden 01109, Germany}\\
\normalsize{$^{5}$University of Texas at Dallas, Dallas, TX 75080, USA}\\
\normalsize{$^{6}$TU Braunschweig, Braunschweig, Germany}\\
\normalsize{$^{7}$Sandia National Laboratories, Albuquerque, NM 87123}\\
\normalsize{$^{\dag}$Equal contribution} \\
\normalsize{$^{*}$To whom correspondence should be addressed} \\
\normalsize{Email: sengupta@psu.edu, kni@nd.edu} 
}

\date{}

\begin{document} 

\maketitle 

\vspace{5mm}
\begin{sciabstract}
Neuromorphic systems seek to replicate the functionalities of biological neural networks to attain significant improvements in performance and efficiency of AI computing platforms. However, these systems have generally remained limited to emulation of simple neurons and synapses; and ignored higher order functionalities enabled by other components of the brain like astrocytes and dendrites. In this work, drawing inspiration from biology, we introduce a compact Double-Gate Ferroelectric Field Effect Transistor (DG-FeFET) cell that can emulate the dynamics of both astrocytes and dendrites within neuromorphic architectures. We demonstrate that with a ferroelectric top gate for synaptic weight programming as in conventional synapses and a non-ferroelectric back gate, the DG-FeFET realizes a synapse with a dynamic gain modulation mechanism. This can be leveraged as an analog for a compact astrocyte-tripartite synapse, as well as enabling dendrite-like gain modulation operations. By employing a fully-depleted silicon-on-insulator (FDSOI) FeFET as our double-gate device, we validate the linear control of the synaptic weight via the back gate terminal (i.e., the gate underneath the buried oxide (BOX) layer) through comprehensive theoretical and experimental studies. We showcase the promise such a tripartite synaptic device holds for numerous important neuromorphic applications, including autonomous self-repair of faulty neuromorphic hardware mediated by astrocytic functionality. Coordinate transformations based on dragonfly prey-interception circuitry models are also demonstrated based on dendritic function emulation by the device. This work paves the way forward for developing truly "brain-like" neuromorphic hardware that go beyond the current dogma focusing only on neurons and synapses.


\end{sciabstract}

\section*{\textcolor{sared}{\large Introduction}}


Current neuromorphic computing systems are synapse and neuron centric. Inputs from the pre-synaptic neurons are weighted and then integrated in post-synaptic neurons to generate outgoing spikes. This model, however, is not completely bio plausible. In fact, most synapses are tripartite – having a third terminal from astrocytes, a type of glial cell. Although glial cells constitute $\sim50\%$ of all human brain cells \cite{moller2007glial}, their inability to generate electrical impulses led many to regard them as mere supportive structures of the brain for a long time. Recent discoveries, however, suggest they not only work as sensing and communication mediums for various brain regions but also play an active role in regulating spiking behavior across multitudes of neurons -- underlying various brain functionalities \cite{allen2009glia, kuga2011large}. Likewise, dendrites, intricate structures that receive input from neighboring neurons and channel them to the soma (neuronal cell body) for summation, are not mere interconnecting wires among biological neurons. In fact, they possess “active” processing capabilities, modulating the input gains across various sensory, motor, and cognitive modalities crucial for object perception, non-linear filtering and spatio-temporal processing \cite{salinas2000gain,takahashi2016active, moore2017dynamics, gasparini2006state}. Fig.\ref{fig:introduction}(a) shows the working principle of dendrites and astrocytes. While astrocytes modulate the neurotransmitters in the synaptic cleft by effectively changing their transmission probability \cite{perea2007astrocytes}, dendrites collate inputs from various neurons and modulate it en route to the soma \cite{major2013active}. From memory formation and improving learning to self-repair and neural phase synchrony, contemporary computational analyses have revealed the efficacy of including astrocytic and dendritic components to established artificial neural network platforms \cite{han2023astromorphic, rastogi2021self, garg2021emulation, wu2018improved, chavlis2025dendrites}. \\

In order to bring these critically needed computational complexity to neuromorphic systems, efficient hardware implementations of dendrites and astrocytes are needed. Efforts thus far to realize astrocytic and dendritic functionality in Complementary Metal Oxide Semiconductor (CMOS) hardware have relied on analog circuits \cite{irizarry2013cmos, irizarry2015astrocyte, lee2016cmos, chance2023shunting, cardwell2023dendritic} or low complexity FPGAs \cite{karimi2018neuromorphic, nazari2015digital, nease2011modeling, george2013low}. Besides being bulky, the focus for these approaches has generally been emulation of computational neuroscience models, rather than computation. In similar vein, advances in emerging post-CMOS technologies to implement compact neuromorphic devices \cite{wang2018experimental, saha2021intrinsic} have largely ignored abstractions of astrocytes or dendrites. While some nascent studies have looked at harnessing dendrites \cite{li2020power, chen2023multi,d2024denram} and astrocytes \cite{garg2021emulation}, they generally employ standalone devices that work in conjunction with other synaptic and/or neuronal devices. These approaches have had limited scopes of application and require significant peripheral circuits that often undermine their feasibility. Compact and efficient realizations of astrocytes and dendrites, especially a device that can be reconfigured to mimic both functionalities in addition to synaptic functionalities in the same device itself,  still remain missing. 

In this work, we propose a novel single transistor solution, i.e., double gate ferroelectric field effect transistor (DG-FeFET), that can emulate the functionalities of both astrocytes and dendrites. Fig.\ref{fig:introduction}(a) shows a fully-depleted silicon-on-insulator (FDSOI) FeFET, that intrinsically realizes a DG-FeFET. A ferroelectric thin film layer is integrated with the top gate (TG) of the DG-FeFET such that the nonvolatile polarization states can store the conventional synapse weight, as has been widely studied in \cite{mulaosmanovic2017synapse, jerry2017ferroelectric, yu2021ferroelectric, saha2021intrinsic}. Beyond the simple synaptic function, the FDSOI FeFET also has a back gate (BG) and a non-ferroelectric insulator, i.e., buried oxide (BOX) layer \cite{jiang2022asymmetric, mulaosmanovic2021ferroelectric, zhao2023powering, zhao2024paving}. The device has a thin Si channel sandwiched between the two gates. The TG and BG both control the channel characteristics and are electrically coupled. Therefore, the synapse weight can be temporarily modulated via the BG, on top of the nonvolatile weight set by the polarization, akin to the astromorphic modulation in biological astrocytes or gain modulation in dendrites. In this way, a single transistor with four terminals can realize a compact astrocyte tripartite synapse or a dendritic arbor. Note that this design can also be applicable for other transistor memory technologies, e.g. Charge trap Flash. However, the superior energy efficiency of ferroelectric HfO\textsubscript{2} based FeFET with its excellent scalability, and CMOS-compatibility \cite{mulaosmanovic2021ferroelectricre} make it a prime candidate for this application. Moreover, with the capability of realizing ferroelectricity in HfO\textsubscript{2} at back-end-of-line (BEOL) compatible thermal budget \cite{aabrar2022beol, datta2024amorphous}, it is possible to integrate monolithically in 3D architectures, ensuring maximal density. Together with FeFET based synapses \cite{mulaosmanovic2017synapse, jerry2017ferroelectric}, FeFET based leaky integrate and fire neurons  \cite{mulaosmanovic2018neuron}, and the DG FeFET based astrocyte and dendrite, the DG-FeFET is a prime platform candidate to realize complex neuromorphic systems.


\begin{figurehere}
   \centering
    \includegraphics[scale=1,width=\textwidth]{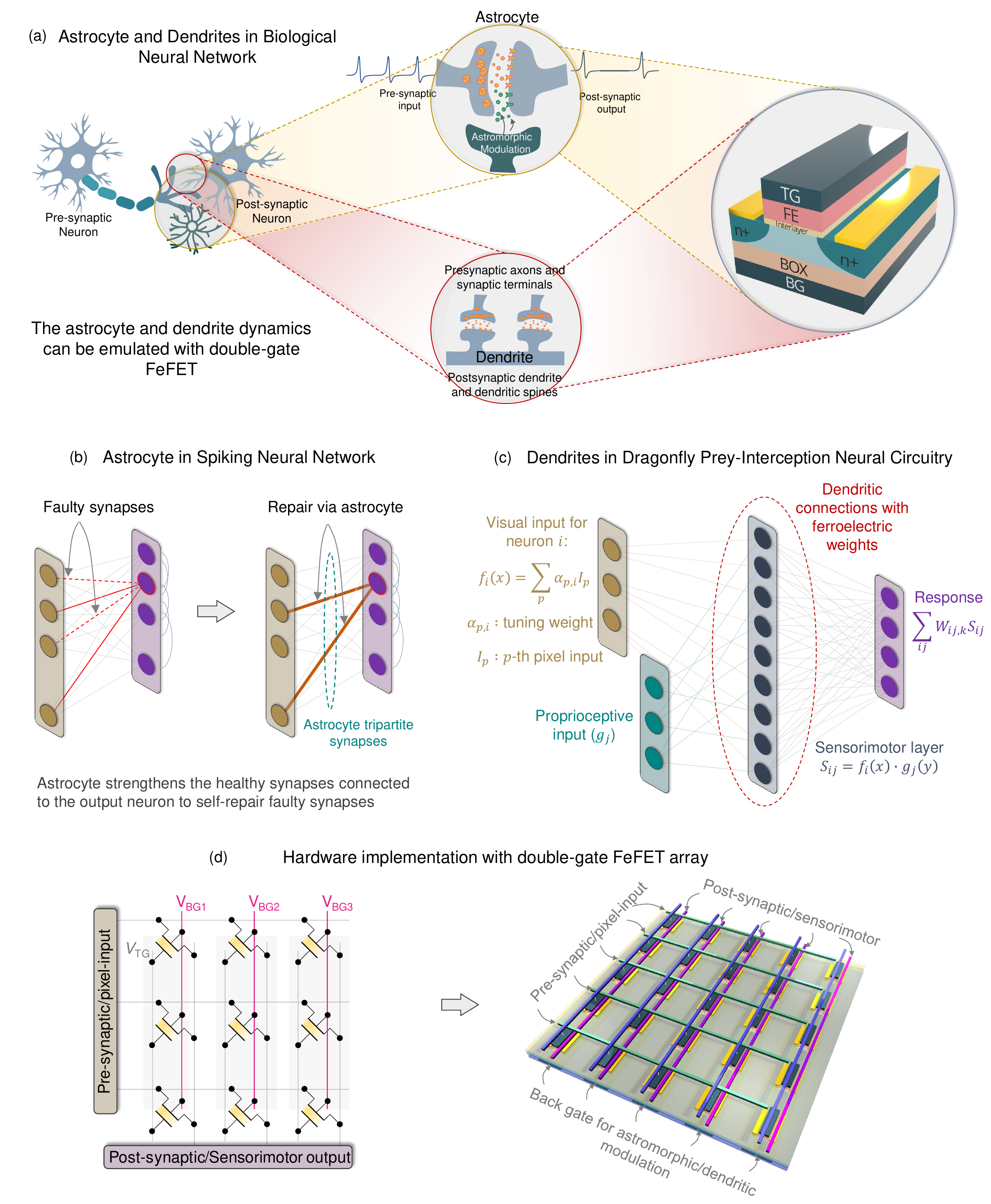}
    \caption{\textit{\textbf{Motivation and potential applications.} (a) Astrocytes and dendrites in biological neural networks perform a plethora of functionalities. We will primarily focus on their ability to modulate the gain of inputs coming in/out from synapses using the Back Gate terminal (BG) of a Double-gate Ferroelectric Field Effect Transistor (DG-FeFET). (b) The DG-FeFET can enable astrocyte tripartite synapses in Spiking Neural Networks, enabling self-repair functionality in hardware. When faults/damage occurs in some synapses, the astrocyte strengthens the healthy synapses to restore similar function levels. (c) Model prey-interception neuronal circuitry of a dragonfly (detailed mathematical model described later in text). Inputs from the visual neuron population are modulated by the proprioceptive neuron population through dendrites enabled by the DG-FeFETs. The final sensorimotor response is a function of both the visual and proprioceptive inputs, which is ultimately used to calculate the final motor response. (d) The DG-FeFET affords efficient hardware implementation with a dual-port FeFET psuedo-crossbar array, where inputs from the pre-synaptic neurons/pixels propagating through the synaptic/tuning weights are modulated by the column-wise connected BGs that perform astrocytic/dendritic modulation. The circuit connections will be discussed in details later for the astrocytic and dendritic applications considered herein.
    }}
    \label{fig:introduction}
\end{figurehere}

The DG-FeFET cell allows a straightforward mapping of the astrocyte and dendrite functionalities to an FeFET array. Fig.\ref{fig:introduction}(b) illustrates a classical two-layer Spiking Neural Network (SNN) with pre-synaptic input neurons and post-synaptic output neurons, but with astrocyte tripartite synapses connected between them. On the other hand, Fig.\ref{fig:introduction}(c) showcases the neural circuitry underlying prey-interception of a dragonfly, where inputs from one population of neurons (encoding visual information) is modulated by dendritic connections coming from another set of neurons (encoding proprioceptive information). In both cases, for an array implementation, as shown in Fig.\ref{fig:introduction}(d), input voltages are fed along the pre-synaptic neuron/pixel input terminals and are modulated by synapse/tuning conductance before being summed up at the post-synaptic neuron/sensorimotor terminals. The back gates of all the synapses along the columns are connected, i.e., the astromorphic/dendritic modulation is mapped into a column-wise BG input to the array. This granularity of BG control per post-synaptic/sensorimotor neuron greatly simplifies the complexity of implementation, saving space for the BG wire routing. Details of the computational model and circuit mapping will be discussed in later sections.
The pseudo-crossbar array can perform the intended matrix-vector multiplication while the cross-point weights can be temporarily modulated via the exerted BG biases. In FDSOI FeFET, the BG is realized with a p-well contact and thus, the implementation of the column-wise BG can also be realized through column-wise p-well contacts and necessary isolation between neighboring p-wells. For the BEOL DG-FeFET implementation, it is possible to directly adopt a metal BG, such that the density of integration is higher than the FDSOI FeFET. Here, it should be noted that although single transistor arrays can be successfully operated \cite{jiang2022feasibility, xiao2022write}, if needed, access transistors can be added to each cell for better management of write disturbance \cite{saha2021intrinsic}.

Such a kernel with an array of astrocyte tripartite synapses or dendritic arbors, if available, could enable multitudes of important applications, such as efficient hardware emulation of biological models such as a dragonfly prey interception model \cite{chance2020interception} as shown in Fig.\ref{fig:introduction}(c) and SNN-based self-repair \cite{navarrete2010endocannabinoids} as shown in Fig.\ref{fig:introduction}(b).
For successful prey interception, a dragonfly must transform visual information in ``eye" or ``camera" coordinates into motor commands expressed in ``body" coordinates. To perform this transformation, the dragonfly prey interception model relies upon a set of basis functions encoded by a layer of gain-modulated units that combine visual information with dragonfly head position information via a dendritic gain control mechanism, which can be supported by the DG-FeFET. For the self-repair application, the astrocyte tripartite synapse enabled by the DG-FeFET can perform self-repair, like in the biological brain, by boosting healthy synapses to compensate for damaged synapses -- ensuring equivalent firing rates. Additionally, the extra terminal for modulating synaptic weights in DG-FeFET offers a new method for homeostasis, a crucial process for neuron activity regulation, as illustrated in Fig.\ref{fig:homeostasis}.

\section*{\textcolor{sared}{\large Double-Gate FeFET for Emulating Astrocyte and
Dendrite Dynamics}}


The DG-FeFET, implemented with FDSOI FeFET, has a top ferroelectric gate that can be used to program the synaptic weight by exploiting partial polarization switching of the ferroelectric layer to access the analog states and the bottom back gate can act as astrocytic/dendritic terminal to modulate the channel conductance. To understand how the BG bias, \textit{V}\textsubscript{BG}, modulates the DG-FeFET, it is necessary to look at the electrostatics and transport properties. Fig.\ref{fig:device_principle}(a) and (b) show the gate stack and equivalent circuit symbols, respectively. It has been well established that the top gate device threshold voltage (\textit{V}\textsubscript{TH}) can be linearly shifted by the \textit{V}\textsubscript{BG} \cite{mulaosmanovic2021ferroelectric, lim1983threshold}, with a linear coefficient $\gamma$\textsubscript{TG}, which is defined as
\begin{equation} \label{Eq.1}
    \gamma_{TG} = \cfrac{\cfrac{C_{CH}*C_{BGOX}}{C_{CH}+C_{BGOX}}}{C_{TGOX}}
\end{equation}
\noindent where the \textit{C}\textsubscript{TGOX}, \textit{C}\textsubscript{CH}, and \textit{C}\textsubscript{BGOX} are the top gate oxide capacitance, the channel capacitance, and the back gate oxide capacitance, respectively, as defined in Fig.\ref{fig:device_principle}(a) and (b). With this linear dependence, the channel conductance, \textit{G}\textsubscript{DS}, sensed at the top gate at a given \textit{V}\textsubscript{BG} can be related with the \textit{G}\textsubscript{DS} at \textit{V}\textsubscript{BG}=0V, as explained in Fig.\ref{fig:device_principle}(c). The relationship can be expressed as
\begin{equation} \label{Eq.2}
    G_{DS}(V_{BG}) = \frac{\mu_n(V_{BG})}{\mu_n(V_{BG}=0)}G_{DS}(V_{BG}=0)+\gamma_{TG}\cdot\mu_n(V_{BG})\cdot C_{TGOX} \cdot V_{BG}
\end{equation}
where the $\mu_n(V_{BG})$ and $\mu_n(V_{BG}=0)$ are the electron mobility at a nonzero and zero \textit{V}\textsubscript{BG}, respectively. One interesting property of FDSOI transistor is that the electron mobility increases with the \textit{V}\textsubscript{BG}. This is because at a higher \textit{V}\textsubscript{BG}, the carrier centroid in the channel will move away from the interface between the top gate dielectric and channel, thus reducing the Coulomb scattering and surface roughness scattering \cite{nier2013multi, al2022impact, han2022back}. Within a reasonable range of \textit{V}\textsubscript{BG}, the mobility can be approximated as a linear function of \textit{V}\textsubscript{BG} \cite{nier2013multi}. Therefore, the \textit{G}\textsubscript{DS} at a nonzero \textit{V}\textsubscript{BG} will be linearly dependent on the \textit{G}\textsubscript{DS}(\textit{V}\textsubscript{BG}=0) with a slope proportional to \textit{V}\textsubscript{BG}. Fig.\ref{fig:device_principle}(d) and (e) illustrates a qualitative picture of \textit{G}\textsubscript{DS} sensed at different \textit{V}\textsubscript{BG}s during synaptic potentiation and depression and the corresponding \textit{G}\textsubscript{DS} dependence on the \textit{G}\textsubscript{DS}(\textit{V}\textsubscript{BG}=0) with varying \textit{V}\textsubscript{BG}. 

\begin{figurehere}
   \centering
    \includegraphics[width=1\textwidth]{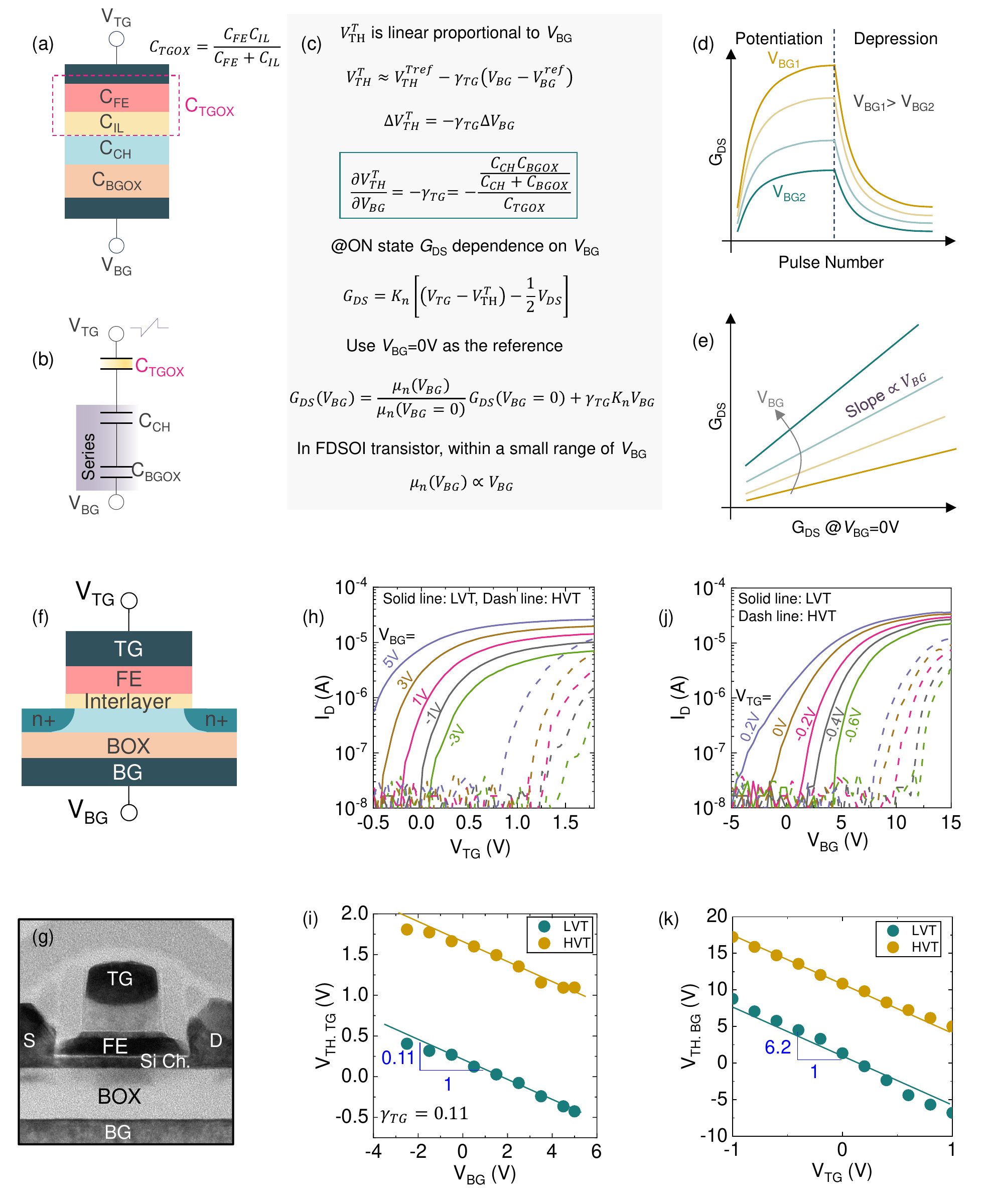}
    \caption{\textit{\textbf{Working principles of double-gate FeFET.}
    (a) The complete gate stack for the DG-FeFET. (b) Equivalent circuit symbols for the gate stack. (c) Explanation for the linear relationship of \textit{G}\textsubscript{DS}(\textit{V}\textsubscript{BG}) with \textit{G}\textsubscript{DS}(\textit{V}\textsubscript{BG} = 0). (d) Qualitatively showing how \textit{G}\textsubscript{DS} varies with potentiation and depression pulses. (e) The qualitative relationship between \textit{G}\textsubscript{DS}(\textit{V}\textsubscript{BG}) and \textit{G}\textsubscript{DS}(\textit{V}\textsubscript{BG}=0V) for various \textit{V}\textsubscript{BG}. The slope is proportional to the applied BG voltage, \textit{V}\textsubscript{BG}. (f) Schematic of the DG-FeFET integrated on 22nm FDSOI technology platform. (g) Transmission Electron Microscope (TEM) view of the device cross-section. (h) \textit{I}\textsubscript{D}-\textit{V}\textsubscript{G} characteristics sensed at TG for different values of \textit{V}\textsubscript{BG}. (i) The threshold voltage as sensed from the TG follows a linear relationship with respect to \textit{V}\textsubscript{BG}. (j) \textit{I}\textsubscript{D}-\textit{V}\textsubscript{G} characteristics sensed at BG for different values of \textit{V}\textsubscript{TG}. (i) The threshold voltage as sensed from the BG also follows a linear relationship with respect to \textit{V}\textsubscript{TG}.        
    }}
    \label{fig:device_principle}
\end{figurehere}



Next, experimental validations are performed to confirm the intended functionalities. Industrial FDSOI FeFETs integrated on 22nm FDSOI technology platform is adopted for experimental testing \cite{dunkel2017fefet} (Fig.\ref{fig:device_principle}(f)). The device's cross section obtained with transmission electron microscopy is shown in Fig.\ref{fig:device_principle}(g). The device
features a $10$nm thick atomic layer deposited doped HfO\textsubscript{2} as the ferroelectric. Detailed process information is described in \cite{dunkel2017fefet}. Fig.\ref{fig:device_principle}(h) and (i) show the \textit{I}\textsubscript{D}-\textit{V}\textsubscript{G} characteristics and corresponding \textit{V}\textsubscript{TH} measured at the top gate, respectively, for the low-\textit{V}\textsubscript{TH} and high-\textit{V}\textsubscript{TH} states sensed at different \textit{V}\textsubscript{BG}s. It clearly shows that \textit{V}\textsubscript{TH} for both states decreases linearly with the \textit{V}\textsubscript{BG} due to electrostatic coupling. The memory window (MW), however, remains constant with the \textit{V}\textsubscript{BG}. Similarly, Fig.\ref{fig:device_principle}(j) and (k) show the \textit{I}\textsubscript{D}-\textit{V}\textsubscript{G} characteristics sensed at the bottom gate and corresponding \textit{V}\textsubscript{TH} at different front gate biases. It also shows that the back gate \textit{V}\textsubscript{TH} is linearly dependent on the front gate bias. An interesting fact is that the memory window sensed at the back gate is much higher than that sensed at the front gate, which is due to the thick gate dielectric at the back gate. This window amplification, however, can not increase the number of available states as it amplifies both the window and the variation of the front gate \cite{chatterjee2022comprehensive}.  

The synaptic operation with astrocyte/dendrite gain modulation is then characterized for the DG-FeFET. Following standard practice for artificial synapses, three pulse schemes are applied: identical pulse trains, pulse width modulation, and pulse amplitude modulation. The schemes are illustrated in Fig.~\ref{fig:synapse}(a), (b), and (c), respectively along with the corresponding gate pulse amplitudes and widths for each scheme. During synaptic potentiation and depression, the source, drain, and back gate are grounded. Conductance measurements are performed by sweeping the \textit{I}\textsubscript{D}-\textit{V}\textsubscript{G} characteristics on the top gate under different \textit{V}\textsubscript{BG} conditions. The corresponding conductance tuning characteristics resemble those reported previously \cite{jerry2017ferroelectric} on FeFET synapses. The identical pulse scheme yields the worst tuning characteristics, where it exhibits very limited number of states. The number of intermediate states can be increased with varying pulse widths as more domains can participate in the switching, but there is a strong asymmetry and nonlinearity between potentiation and depression. In contrast, the pulse amplitude modulation exhibits the best characteristics, where it shows an almost linear and symmetric potentiation and depression behavior. Aside from confirming the well-established synaptic behavior in FeFETs, these results also validate that the synaptic weight can be modulated by the \textit{V}\textsubscript{BG}, where the higher the \textit{V}\textsubscript{BG}, the higher the conductance, thus showing the desired astrocyte/dendrite function. Such weight programming and astrocyte/dendrite modulation are reproducible, as seen in 10 times repeated measurement of the pulse width modulation scheme shown in Fig.\ref{fig:synapse}(d) and the pulse amplitude modulation scheme shown in Fig.\ref{fig:synapse}(e).

In plotting the measured conductance at different \textit{V}\textsubscript{BG}s with respect to that at \textit{V}\textsubscript{BG}=0V, as shown in Fig.\ref{fig:synapse}(f), a linear dependence is presented, consistent with Eq.\ref{Eq.2}. In addition, the extracted slope of the linear fitting, as shown in Fig.\ref{fig:synapse}(g), is also linearly dependent on the \textit{V}\textsubscript{BG}, again confirming the theoretical analysis shown in Eq.\ref{Eq.2}. It also indicates that within a reasonable range of \textit{V}\textsubscript{BG}, the mobility is linearly increasing with the \textit{V}\textsubscript{BG}. Therefore, these results establish that the device conductance is a product between the \textit{V}\textsubscript{BG} and the nonvolatile conductance at \textit{V}\textsubscript{BG}=0V, which emulates a function of astrocyte/dendrite-modulated synapse. The weight modulation is also highly robust and reproducible, therefore validating the astrocyte/dendrite functionality. Next, the state stability is measured. First, the switching dynamics of the FDSOI FeFET is characterized, as presented in supplementary Fig.\ref{fig:dynamics}. It shows that there is a large basin around 0V, indicating it is difficult to flip the polarization states around 0V, thus good for retention. This has been verified by measuring the retention for the intermediate states at room temperature. As shown in Fig.\ref{fig:synapse}(h), there is negligible degradation for all the states, showing the nonvolatility. Therefore, all the aforementioned combined theoretical and experimental analysis validate the capability of DG-FeFET in realizing the 
astrocyte/dendrite-modulated synapse.

\begin{figurehere}
   \centering
    \includegraphics[width=1\textwidth]{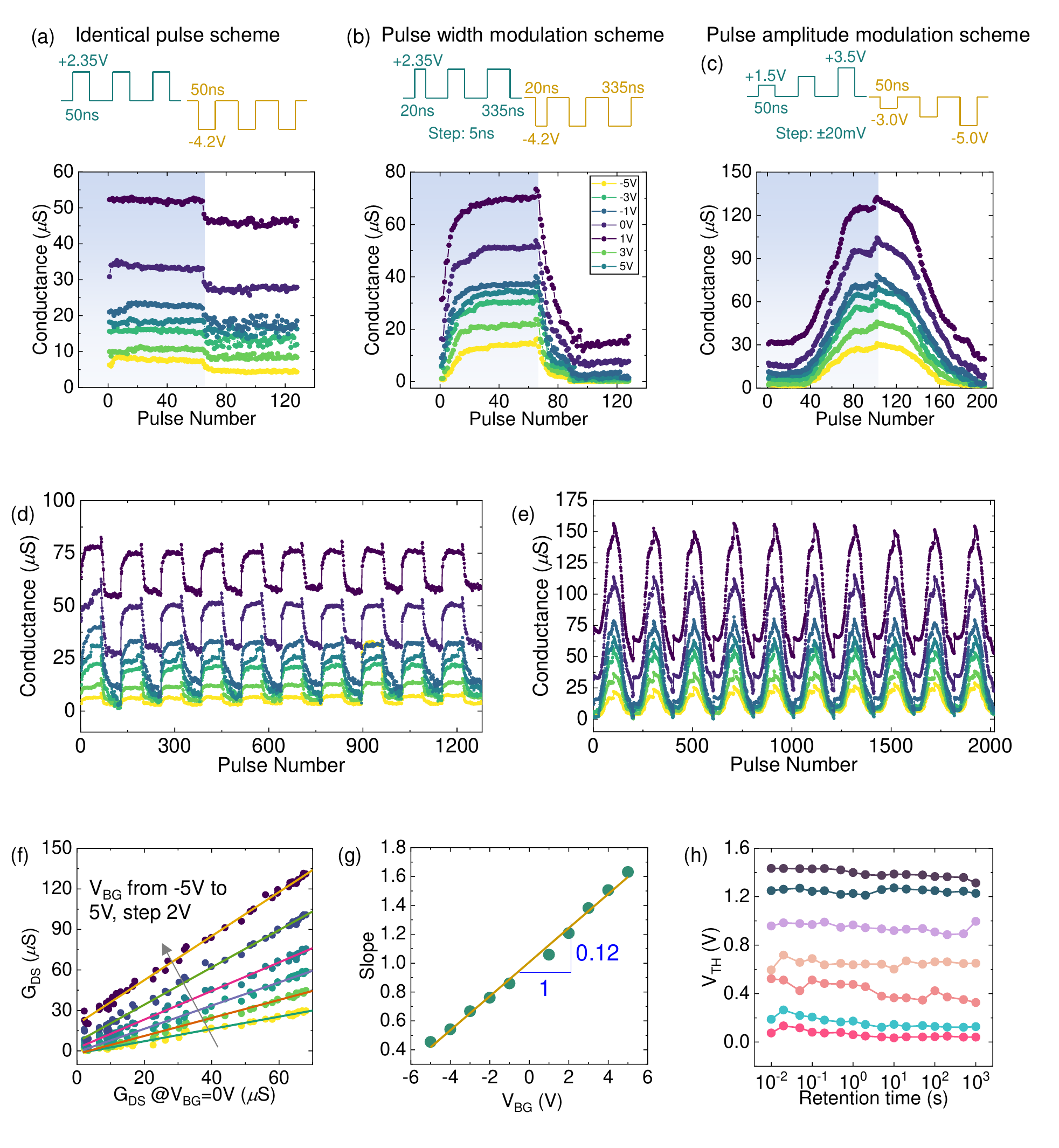}
    \caption{\textit{\textbf{Synaptic plasticity with astrocyte/dendrite modulation for the DG-FeFET.} Three different pulsing schemes applied to the TG with different BG voltages: (a) identical pulses, (b) pulse width modulation, (c) pulse amplitude modulation. Pulse amplitude modulation shows the best linearity of programming for both potentiation and depression. (d) The weight programming with the astrocyte/dendrite modulation shows reproducibility over 10 cycles for the pulse width modulation scheme and (e) pulse amplitude modulation scheme. (f) Linear relationship between $G_{DS}$ for various $V_{BG}$ vs $G_{DS} @ V_{BG}=0V$. (g) The slope of the fitted lines in (f) increase linearly with $V_{BG}$. (h) The DG-FeFET device states show excellent retention with negligible degradation for upto $10^3$s, showing the nonvolatile nature of all the states.
    }}
    \label{fig:synapse}
\end{figurehere}


\section*{\textcolor{sared}{\large Application of Astrocytes and Dendrites enabled by DG-FeFET}}

We experimentally evaluate the capability of a DG-FeFET array to perform the multiply-accumulation operation. Fig.\ref{fig:application}(a) shows a 7$\times$8 array that is used for demonstration. The column current is measured when activating different number of cells in that array. Fig.\ref{fig:application}(b) shows the swept current as a function of top gate bias for all the drain lines (i.e., 8 lines) in the array. With the read bias at 2V, the column current shows a linear trend with the number of activated cells, as shown in Fig.\ref{fig:application}(c). With the crossbar macro introduced in Fig.\ref{fig:introduction}(d), several applications of the astrocytic/dendritic operation enabled by the DG-FeFET is considered.

\noindent
\textbf{Self-Repair with Astrocyte Tripartite Synapse}: Self-repair functionality, as introduced in Fig.\ref{fig:introduction}(b), has been attributed to astrocytes in the biological brain. The DG-FeFET's BG offers a hardware solution to enable this directly in hardware.   
Fig.\ref{fig:application}(d) illustrates the phenomenon of self-repair in time domain. When faults happen in some of the synapses, i.e., causing no contribution to the output neurons from corresponding inputs, the accuracy/performance of the network degrades. To mitigate that, the 'healthy' astrocyte tripartite synapses can be boosted with the \textit{V}\textsubscript{BG} to recover similar output levels and restore the accuracy. To test the self-repair capabilities on real-world data, we train and test the network on the MNIST handwritten digit recognition dataset \cite{lecun2010mnist}. Details regarding the network and simulation are in the Supplementary section. We obtain a baseline test accuracy of $88.18\%$ on the MNIST test set. Fig. \ref{fig:application}(e) shows the pristine learnt weights. Note, the accuracy of the network can be increased even further by changing the total number of output neurons \cite{diehl2015unsupervised}. Here, we are mainly concerned about the hardware’s self-repair capabilities and thus keep the network size relatively small to reduce the total simulation times. After the initial training, we introduce stuck-at-zero faults (i.e., short-circuits) to the learnt weight patterns (Fig. \ref{fig:application}(f)). The network accuracy drops significantly as the percentage of faults increases as indicated by Table \ref{tab:fault_injection}.

To recuperate the accuracy back, we want to employ the astrocyte-like repair mechanism to our “healthy” synapses (Fig. \ref{fig:application}(e-g)). Previous studies on astrocyte computational models realize self-repair by retraining the network with faults in place with an additional astrocytic self-repair scaling factor in the learning rule. This factor dynamically encodes the degree by which the neuron firing rate deviates from the original firing frequency, thus ensuring a similar firing activity after retraining \cite{han2023astromorphic, rastogi2021self}. Obviously, this does not map directly to device dynamics and thus, implementing it on-chip at each cross-point requires significant complexity in the peripherals. Propitiously, the back-gate allows us a much simpler way to implement the factor, while preserving the advantage of using the original device dynamics as shown in eq. \ref{Eq_FeFET_STDP}. By boosting the conductance with the back-gate voltage, $V_{BG}$, the network sees the effective weight, $W_{eff}$ in lieu of the originally programmed post-synaptic weights, $W_{post}$ and modulates the incoming spikes. We can write:\\
\vspace{-7mm}
\begin{equation}
    \label{Eq_I_neuron_self}
    W_{eff} = k*W_{post}
\end{equation}
\vspace{-10mm}

\noindent Here, $k$ is a function of the applied back gate voltage to each column and is henceforth referred as the degree of self-repair. The degree of self-repair needed for each column of weights is adapted as the network is retrained. Further details of the self-repair functionality can be found in the Supplementary Section 'Self-repair Mechanism of the Network'. Table \ref{tab:fault_injection} shows that our self-repair scheme can greatly recover its lost accuracy. Effectiveness of the self-repair process becomes obvious as the number of faults in the network increases. Modifying the weights without having such a back-gate control can not only result in no recovery but in some cases can worsen the network performance after the retraining phase (Table \ref{tab:fault_injection}) – further indicating the efficacy of the back-gate voltage. In addition to self-repair, astrocytes have also been shown to possess homeostatic functionality, playing a critical role in neural activity regularization and ensuring learning \cite{han2023astrocyte}. Interestingly, the DG-FeFET array can be also used to intrinsically implement homeostasis directly in the crossbar. Further details are discussed in the Supplementary Text.

\noindent
\textbf{Dragonfly Prey-Interception Model with Dendritic Gain Modulation:} A DG-FeFET array can be used to implement a model of dragonfly prey-interception circuitry \cite{chance2020interception}, illustrated in Fig.\ref{fig:introduction}(c) and described in more detail in Supplemental Information. The dragonfly circuit model relies upon gain modulation to perform a coordinate transform operation. Briefly, visual input (illustrated in yellow in Fig.\ref{fig:introduction}(c)) is modeled as the input, $f(x)$, from a population of neurons, where $f_i(x)$ is the response of neuron $i$ to $x$, the position of the prey's image on the dragonfly's eye. Similarly, $g(y)$ is proprioceptive input from a population of neurons for which $g_j(y)$ is the response of neuron $j$ encoding the desired position, $y$, of the prey's image on the dragonfly's eye for correctly intercepting the prey. 
The visual input is modeled as sensory input, $I_p(x)$, directly from the ommatidia (analogous to pixels) of the dragonfly eye mapped to the drain input of an array of DG-FeFET devices. We model the input from visual input neuron $i$, $f_{i}(x)$, by programming a tuning weight,
\begin{eqnarray}
    \alpha_{p,i} = \kappa \exp\left( -\frac{(x_{p}-a_{i})^2}{2\sigma_{r}^2}\right),
\end{eqnarray}
using the top gate input, where $x_{p}$ is the visual location encoded by ommatidium $p$, $a_{i}$ is the ``preferred'' visual location of visual input neuron $i$, and $\kappa$ is a synaptic scaling parameter. Input from proprioceptive neuron $j$, $g_{j}(y)$ (lying in the range of [0, 1] \cite{chance2020interception}) is mapped to the back gate using the slope values given in Fig.\ref{fig:synapse}(g). This mapping introduces an offset into the values of $g$, which can be easily corrected by introducing additional steps into the turn decoding algorithm.The summed output from the DG-FeFET array is proportional to $S_{ij}(x,y)=f_{i}(x)g_{j}(y) = \sum_p{I_p(x)\alpha_{p,i}}g_{j}(y)$. We envision this array of DG-FeFET devices as modeling the dendritic arbor of one sensorimotor neuron in Fig.\ref{fig:application}(h). It is worth mentioning here that $I_p(x)$ and $g_j(y)$ are dynamically changing inputs while $\alpha_{p,i}$ is static, thereby leading us to program $\alpha_{p,i}$ using the FeFET top-gate.

Fig.\ref{fig:application}(i) are the responses of one sensorimotor neuron, modeled by one such array of DG-FeFET devices. Each curve is the sensorimotor neuron response as the prey travels along a trajectory orthogonal to the orientation of the dragonfly. To produce the different curves, we vary the level of back-gate input, demonstrating the multiplicative effect of the back gate. The sensorimotor layer is configured such that all possible combinations of visual and proprioceptive inputs are represented. While the visual input is in eye coordinates (relative to the reference frame of the eye), this configuration of input combinations may be thought of as forming a set of basis functions \cite{pouget1997spatial,pouget2000computational} from which motor outputs may be calculated for alternate frames of reference (e.g., body-coordinates). This sensorimotor transformation is achieved by summing linear combinations of the basis functions, or sensorimotor neuron outputs. In the dragonfly application, the turns required for interception are represented in the responses of the motor neuron population. The response of an individual motor neuron, $R_k$, is equal to $\beta\sum_{i,j}W_{ijk}S_{ij}(x,y)$ (see Supplemental Information for a description of how the weights are calculated), where $\beta$ controls the overall level of activity of the motor neurons. This weighted dot-product can be implemented using a standard FeFET array. Fig.\ref{fig:application}(j) illustrates an example trajectory (green symbols) of the dragonfly model from \cite{chance2020interception} intercepting a moving prey (red symbols). For simplicity, the model and prey are constrained to move only within a two-dimensional plane, significantly decreasing the size of the model. The purple symbols represent the trajectory of the DG-FeFET array based dragonfly prey-interception model.

\begin{figurehere}
   \centering
    \includegraphics[width=0.95\textwidth]{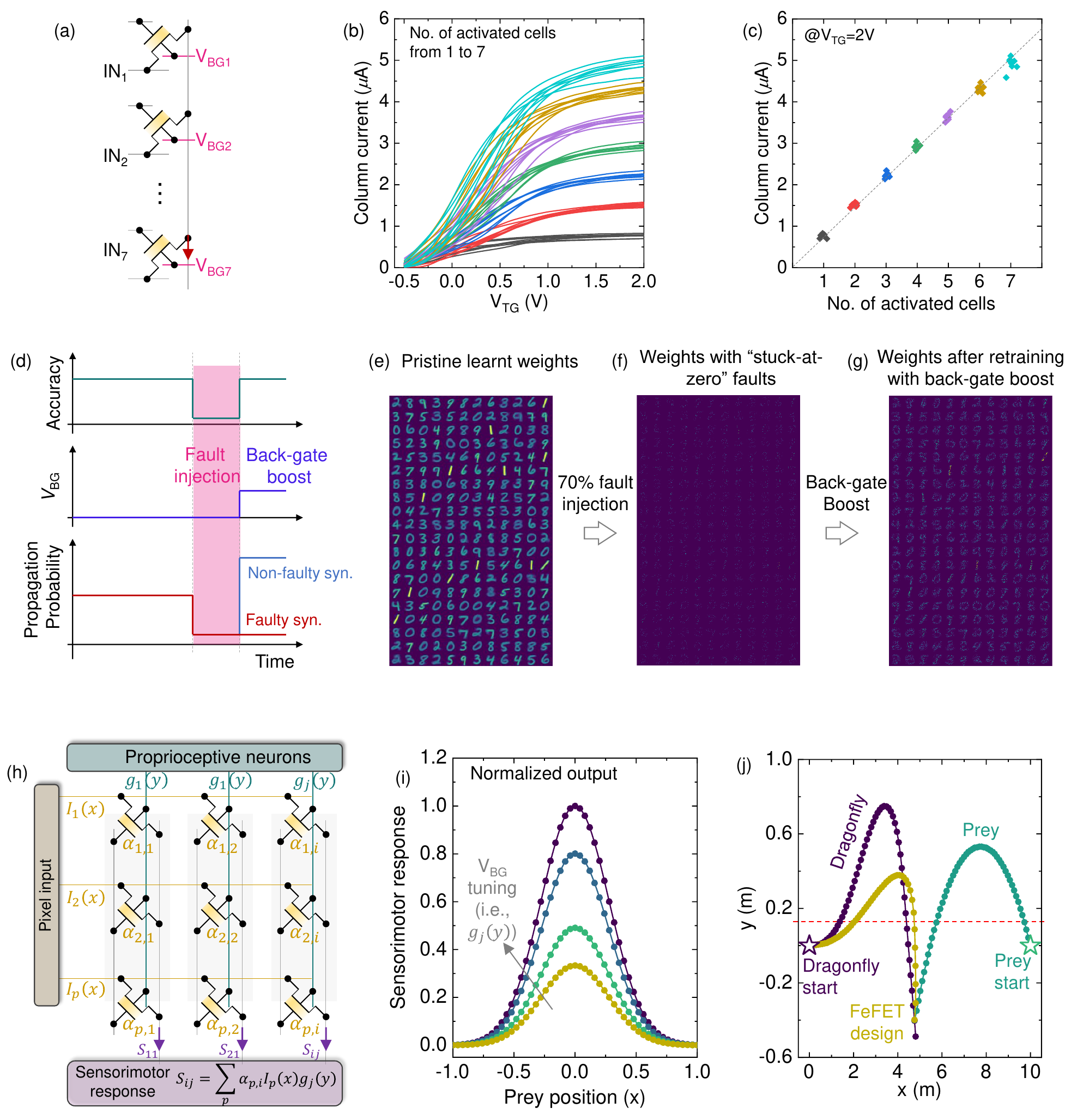}
    \caption{\textit{\textbf{Application of Astrocyte and Dendrite Dynamics enabled by DG-FeFET.} (a) Schematic of the experimental 7$\times$8 array of DG-FeFET (1 column is shown due to independency among columns). (b) Column current as a function of top gate bias for all the drain lines (i.e., 8 lines) in the array. (c) With the read bias with $V_{TG}$ at 2V, the column current shows a linear trend with the number of activated cells. (d) The phenomenon of self-repair in time domain. When faults occur in some synapses, the back-gate voltage enables us to dynamically increase the effective weights of non-faulty/healthy synapses. 
    (e) The pristine learnt weights after training. (f) The weights after the introduction of ``stuck-at-zero" or grounding faults. (g) Effective weights after re-training with BG boost enabled. (h) Recreating the dragonfly prey-interception circuit model with DG FeFET devices where the implementation of the sensorimotor layer is shown. The total number of required columns will be the product of the neuron count in the visual and proprioceptive layers. Input pixels $I_p$ are applied at the drain terminals. The tuning weights $\alpha_{p,i}$ are programmed using the top-gate and the $g_j(y)$ values are applied at the back-gate terminal.
    (i) Example responses of a sensorimotor neuron for different prey locations and different levels of gain modulation via the back gate. Responses are normalized to the maximum response.
    (j) An example trajectory (green symbols) of the dragonfly model intercepting a moving prey (red symbols). With the use of the FeFET dynamics, the dragonfly is able to quickly adjust and incept its prey.   
    }}
    \label{fig:application}
\end{figurehere}
\section*{\large Conclusion}

In this work, we exploit a DG FeFET to implement a compact synapse with dendritic and astrocytic functionalities, whose hardware implementation have long remained elusive. We theoretically and experimentally established the linear dependence of synaptic weight on the astrocyte/dendrite terminal, i.e., BG. Through combined device-to-algorithm co-design, we demonstrate interesting applications, including self-repair, dragonfly prey-interception and homeostasis, that are enabled by the DG FeFET. These results show that DG FeFET, as an efficient hardware implementation of the synapse with astrocytes/dendrites, can augment existing neuromorphic hardware with new functionalities.

\section*{\large Data availability}
All data that support the findings of this study are included in the article and the Supplementary Information file. These data are available from the corresponding author upon request.

\section*{\large Code availability}
All the codes that support the findings of this study are available from the corresponding author upon request.

\bibliography{ref.bib}

\bibliographystyle{naturemag}

\section*{\large Acknowledgements}

This characterization of the DG FeFET is supported by the U.S. Department of Energy, Office of Science, Office of Basic Energy Sciences Energy Frontier Research Centers program under Award Number DESC0021118 (to K.N.). It is also partially supported by SUPREME center, one of the SRC/DARPA JUMP 2.0 centers, and the NSF 2346953. This testing chips are partially funded by the European Union within "NextGeneration EU", by the Federal Ministry for Economic Affairs and Climate Action (BMWK) on the basis of a decision by the German Bundestag and by the State of Saxony with tax revenues based on the budget approved by the members of the Saxon State Parliament in the framework of “Important Project of Common European Interest - Microelectronics and Communication Technologies", under the project name “EUROFOUNDRY”. The application of DG-FeFET was sponsored primarily by the Army Research Office and was accomplished under Grant Number W911NF-24-1-0127. The views and conclusions contained in this document are those of the authors and should not be interpreted as representing the official policies, either expressed or implied, of the Army Research Office or the U.S. Government. The U.S. Government is authorized to reproduce and distribute reprints for Government purposes notwithstanding any copyright notation herein. It was also supported partially by the National Science Foundation under award No. BCS \#2031632, EFRI BRAID \#2318101, CNS \#2137259 - Center for Advanced Electronics through Machine Learning (CAEML) and its industry members.

\section*{\large Author contributions}

A.S. and K.N. proposed and supervised the project. Z.J., Z.Z. J.D., F.M., and T.K. conducted measurements on single cell and array. A.N.M.N.I., Z.H., S.G.C., and F.C. performed application-level analysis. H.M., S.D., and S.B. processed and supplied the testing wafer. S.D. and V.N. advised on the experiment design and data analysis. All authors contributed to write up of the manuscript.

\section*{\large Competing interests}
The authors declare that they have no competing interests.

\newpage

\renewcommand{\thefigure}{S\arabic{figure}}
\renewcommand{\thetable}{S\arabic{table}}
\renewcommand{\theequation}{S\arabic{equation}}
\setcounter{figure}{0}
\setcounter{table}{0}
\setcounter{equation}{0}

\newpage
\begin{center}
\title{\textbf{\Large Supplementary Materials}}
\end{center}

\vspace{-4ex}

\section*{\large Experimental Details}
The electrical characterization was conducted using a measurement setup comprising a PXIe system by NI. A custom switch matrix controlled by NI PXIe-6570 Pin Parametric Measurement Units (PPMU) was used to access each contact of the crossbar array. This setup enables the selection of NI PXIe-4143 Source Measure Units (SMU) for the write operation and PPMU usage for control signals. The crossbar contains a 7$\times$8 array of FeFETs with dimensions of 500$\times$500 nm$^2$. Each bitline is connected to a transistor that acts as a resistive element to limit the On-Current (ION). The switch matrix connects to a probe card that interfaces with the crossbar structure. 
For the accumulate operation, all FeFETs in the crossbar are initially erased by applying a voltage of -5V to the wordlines connected to the gates for 500 ns. To write the FeFETs to LVT, a program voltage of 4.5V is applied for 500 ns. For the read operation, a sweep voltage in the range of -0.5V to 2.0V is applied to the gates of the selected FeFETs, while the unselected FeFETs are disabled by applying a pull-off voltage of -1V to their gates. Additionally, a drain voltage of 50 mV is applied during the read operation. To demonstrate the linear accumulation capability, a readout of one to seven FeFETs in parallel is performed. The sum current is limited by a transistor connected to the drain line, with ION being controllable via the transistor's gate voltage. 

\newpage
\section*{\large FeFET Characteristics}

The switching dynamics of HfO\textsubscript{2} FeFET integrated on 22nm FDSOI platform is shown in Fig.\ref{fig:dynamics}. It shows the device memory window under different write pulse amplitudes and pulse widths. It shows fast write operation ($<$20ns at 4V write and slow switching at 0V during retention.)

\begin{figurehere}
   \centering
    \includegraphics[scale=1,width=0.5\textwidth]{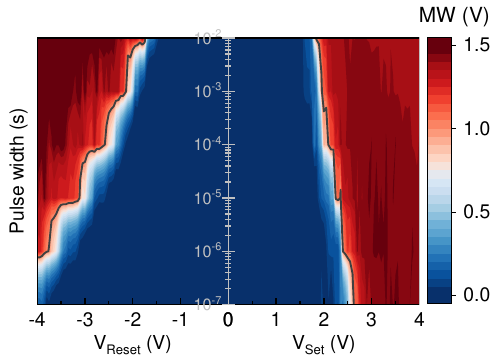}
    \caption{Switching dynamics of HfO\textsubscript{2} FDSOI FeFET. }
    \label{fig:dynamics}
\end{figurehere}

\newpage
\section*{\large Network Simulation}

We train a neural network with 784 input neurons and 400 output neurons in an unsupervised manner with our FeFET devices as synapses. Previously, it has been showed that multi-state FeFET device dynamics emulates a weight-dependent or multiplicative version of the Spike-timing dependent plasticity (STDP) learning \cite{bi2001synaptic}, which introduces dependence on the device state and ultimately results in faster training convergence \cite{saha2021intrinsic}. The synaptic weight update, $\Delta w$ for such devices can be written as:
\begin{equation}
    \label{Eq_FeFET_STDP}
    \Delta w = 
        \begin{cases} 
        (1-w)^{-\mu_+}\times A_+ \times e^{e^{-\Delta t / \tau_+}}, & \Delta t \geq 0 \\
        w^{\mu_-} \times A_- \times e^{e^{-\Delta t / \tau_-}}, & t < 0
        \end{cases}
\end{equation}

\noindent Here, $\Delta w$ is the programmed weight of the FeFET device (corresponds to a specific conductance value), $A_{+/-}$ are the pre- and post-synaptic learning rates, $\mu_{+/-}$ and $\tau_{+/-}$ are experimentally calibrated device simulation parameters obtained from \cite{saha2021intrinsic}.

The output neurons are each connected to all the other output neurons using fixed weight inhibitory connections. This ensures each neuron learns at a time. In order to ensure no single neuron dominates the learning procedure, an adaptive thresholding scheme is adopted. For the homeostasis simulation, this was turned off and instead the BG gain was used for ensuring homeostasis. The network was implemented using the PyTorch based BindsNet framework \cite{hazan2018bindsnet}. Table \ref{tab:net_train_params} lists the network training and retraining parameters.

\begin{table}[h]
\caption{Network Training Parameters}
\vspace{2ex}
    \centering  
    \footnotesize
    \begin{tabular}{p{0.4\linewidth}  p{0.2\linewidth}}
    \hline
    \textbf{Parameters} & \textbf{Values}\\
    \hline
Simulation time duration per image& $100ms$\\
Simulation time-step size& $1ms$\\
Membrane potential decay time const.& $100ms$\\
Refractory period& $5ms$\\
Resting membrane potential& $-65mV$\\
Reset membrane potential& $-60mV$\\
Threshold membrane potential& $-52mV$\\
Adaptive threshold increment& $0.05mV$\\
Adaptive threshold decay time const.& $10^7 ms$\\
Batch size& $16$\\
Number of Training epochs& $2$\\
Number of Retraining epochs& $1$\\
Spike trace decay time const.& $20ms$\\
Maximum input spike rate& $128Hz$\\
Inhibitory synaptic weight& $-120$\\
Post-synaptic learning rate& $10^{-2}$\\
Pre-synaptic learning rate& $10^{-4}$\\
    \hline
    \end{tabular}
    
    \label{tab:net_train_params}
\end{table}

\clearpage
\newpage
\section*{\large Self-repair Mechanism of the Network} \label{sec:sup_repair}

After applying the back-gate voltage $V_{BG,i}$, the network weights, $w_{post,i}$ effectively become $w_{eff,i}$. This modulates the incoming spikes which are then integrated along the columns to generate output spikes:
\begin{equation}
    \label{Eq_I_neuron_self}
    I_{neuron} = \sum w_{eff,i} * V_{spike,i} = \sum k_iw_{post,i}*V_{spike,i}
\end{equation}
Here, $i$ is the number of training instance, $k_i$ is the degree of self-repair and is a function of $V_{BG,i}$. $V_{spike,i}$ is the incoming voltage spikes from the pre-synaptic neurons. The initial scaling factor and in turn, the initial back-gate voltage, $V_{BG,0}$ is proportional to the ratio of the total sum of weights prior to fault, $w_{prior}$ (can be obtained from software) and the total sum of weights after faults are introduced, $w_{post,0}$ (i.e., the network embedded in hardware, can be evaluated by applying an all-one voltage along the crossbar rows).

During retraining, based on the timing differences of the incoming and outgoing spikes, the appropriate weight updates, $\Delta w_{i+1}$ are calculated according to Eq. \ref{Eq_FeFET_STDP}, and the corresponding set pulses are applied. During the weight update, which is dependent on the current weight/device’s programmed state, i.e., $w_{post,i}$, the back-gate voltage is set to zero. The updated weights ($w_{post, i+1}$) results in a corresponding updated $V_{BG,i+1}$ ($\propto \sum w_{prior}/\sum w_{post,i+1}$) and $k_{i+1}$. This gives us the new effective weight and enables the network’s self-repair functionality:
\begin{equation}
    \label{Eq_w_eff}
    w_{eff, i+1} = k_{i+1}w_{i+1}= k_{i+1}(w_{post,i}+\Delta w_{i+1})
\end{equation}

\noindent Here, $\Delta w$ is calculated using Eq.\ref{Eq_FeFET_STDP}. It is worth mentioning here that we consider a simplified implementation of astromorphic self-repair where the neuron firing frequency deviation indicator is considered to be proportional to the ratio of sum of weights of the neuron before and after fault. More detailed dynamics can be considered \cite{han2023astromorphic}. Table \ref{tab:fault_injection} shows the accuracy comparison before and after fault injection. 

\vspace{10mm}
\begin{table}[h]
\caption{Accuracy comparison before and after fault injection, and retraining on the MNIST test data set with and without $V_{BG}$ self-repair.}
\vspace{3ex}
    \centering  
    \footnotesize
    \begin{tabular}{p{0.1\linewidth}  p{0.09\linewidth} p{0.1\linewidth} p{0.2\linewidth} p{0.2\linewidth}}
        \hline
        \centering \textbf{Before Fault Injection} & \centering \textbf{Fault Injection} & \centering \textbf{After Fault Injection} & \centering \textbf{Retraining with $V_{BG}$ self-repair (improvement)} & \textbf{Retraining with no $V_{BG}$ self-repair}\\
        
        \hline
        
        \centering Baseline, & \centering 60\% & \centering 77.63\% & \centering 82.12\% (+4.49\%) & 31.56\% \\
        
        \centering 88.18\% & \centering 70\% & \centering 67.12\% & \centering 78.42\% (+11.30\%) & 33.03\% \\ 
        
        & \centering 80\% & \centering 34.03\% & \centering 76.39\% (+42.36\%) & 45.92\% \\ 
        
        & \centering 90\% & \centering 9.84\% & \centering 70.39\% (+60.55\%) & 9.8\% \\
        
        \hline
    \end{tabular}
    
    \label{tab:fault_injection}
\end{table}

\clearpage
\newpage

\section*{\large Homeostasis via BG Tuning}

\noindent
The BG control enables us to implement homeostasis in the crossbar array itself by controlling the effective conductance of an entire column of synaptic devices. In SNNs, homeostasis is essential for learning across all neurons, as it ensures neurons that have fired require a greater input than before to fire again. Without it, only a few neurons dominate, and the network is not able to learn the desired patterns (Fig.\ref{fig:homeostasis}(c,d)). As illustrated in Fig.\ref{fig:homeostasis}(a), homeostasis is generally enabled in neural networks by increasing the threshold required for a neuron to spike after it fires. In hardware implementation, such regularization is supplemented on top of the neuronal circuitry – adding sizable overhead \cite{rajendran2012specifications, indiveri2003low} (Fig.\ref{fig:homeostasis}(b)). The BG control of the DG-FeFET affords us a much more elegant solution (Fig.\ref{fig:homeostasis}(e),(f)). 

The back-gate voltage, $V_{BG}$ can effectively modulate the output of the crossbar going to the neuron devices as the output is a dot product of the input voltage and device conductances. Thus, by ramping down the back-gate potential of a particular column whenever the neuron associated with it fires, we can achieve homeostasis. We substantiate our claim by integrating these synaptic dynamics in an SNN implementation. We train a network with a similar architecture as the self-repair case before on the MNIST dataset with $100/225$ output neurons. As the training will take place in our FeFET devices, we utilize a weight-dependent learning rule for the STDP, as seen for such devices \cite{saha2021intrinsic}. After 2 epochs of training, we find that the network with back-gate modulation is able to induce homeostasis (Fig.\ref{fig:homeostasis}(g,h)), attaining an accuracy of $80.63\%$ for 100 output neurons and $85.74\%$ for 225 output neurons on the test dataset -- on par with spiking networks of similar sizes \cite{dutta2019biologically, diehl2015unsupervised}.

\begin{figurehere}
   \centering
    \includegraphics[scale=1,width=1\textwidth]{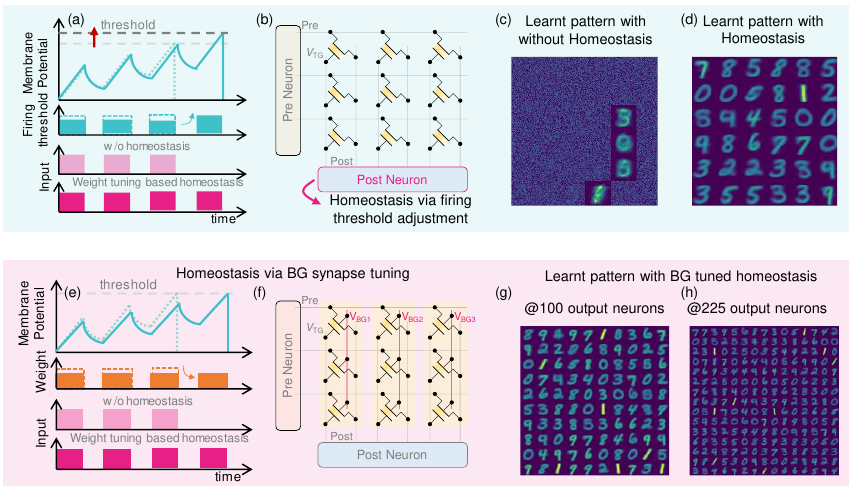}
    \caption{(a) Traditional implementation of homeostasis requires the increase of the neuron threshold potential, i.e., the potential at which the neuron spikes and resets. (b) Conventional cross-bar array and implementation of homeostasis through output neuron threshold adjustment. (c) Without homeostasis, only a few neurons potentiate and try to learn all the patterns. (d) With conventional homeostasis, correct patterns can be learnt. (e) An alternative approach to implement homeostasis and (f) the corresponding array structure using the back-gate voltage, instead of increasing the threshold. The effective weights can be modulated and the increase in the membrane potential can be reduced, thus realizing the same effect. (g-h) The final learnt weights of the network with back-gate enabled homeostasis. 
    }
    \label{fig:homeostasis}
\end{figurehere}

\clearpage
\newpage
\section*{\large Dragonfly Prey-Interception Circuitry Model} \label{sec:sup_dragonfly}

Here, we provide a brief description of a mathematical model of the dragonfly prey-interception circuitry from \cite{chance2020interception}.  As noted in the main text, we first reduce the size of the model by restricting the dragonfly and its prey to move in a two-dimensional plane.  This simplification allowed us to focus on leveraging FeFET devices to emulate the functionalities of individual neurons as predicted by the dragonfly-circuit model.  Inputs from two populations of neurons, visual-input (yellow units in Fig.\ref{fig:introduction}(c) and the proprioceptive input (teal units in Figure Fig.\ref{fig:introduction}(c) of the main manuscript) are combined in the sensorimotor layer (dark blue) such that each sensorimotor neuron, $S_{ij}$, multiplies input $f_{i}(x)$ from visual input neuron $i$ with input $g_{j}(y)$ from proprioceptive input neuron $j$. 

The input received from visual neuron $i$ is: 
\begin{eqnarray}
    f_{i}(x) = \exp\left( -\frac{(x-a_{i})^2}{2\sigma_{r}^2}\right).
\end{eqnarray}
Here $x$ is the position of the prey's image on the dragonfly's eye (described in Cartesian coordinates), $a_i$ is the prey-image position that elicits the strongest input from visual neuron $i$, and $\sigma_{r}$ determines the width of the tuning curves of the visual input neurons.  For the simulated engagement in Fig.\ref{fig:application}(h-j), $x$ is updated every time step as the prey's movements are independent of the dragonfly's. 

The input received from proprioceptive neuron $j$ is:
 \begin{eqnarray}
    g_{j}(y) = \exp \left( \frac{(y-b_{j})^2}{2\sigma_{g}^2} \right),
\end{eqnarray}
where $y$ is the ``desired'' position of the prey's image (described in Cartesian coordinates relative to the dragonfly's eye). As with visual input neurons, $b_{j}$ is the desired position that elicits the strongest input from proprioceptive neuron $j$, and $\sigma_{g}$ determines the selectivity of the proprioceptive neurons.  As the dragonfly turns, $y$ is adjusted to compensate for the effect of the dragonfly's own turning on the prey-image position. We refer to this population of neurons as ``proprioceptive" because this input could be replaced with proprioceptive input describing the dragonfly's head position relative to its body as the dragonfly turns its head to track the prey during flight (see \cite{plunkett2023modeling}).

As mentioned above, individual sensorimotor neurons multiply input from one visual neuron and one proprioceptive neuron such that all possible pairwise combinations of visual and proprioceptive neurons are represented by the sensorimotor populations.  We therefore identify sensorimotor neurons by the visual and proprioceptive neurons from which they receive input -- the activity of sensorimotor neuron $S_{ij}$ is therefore:
\begin{eqnarray}
S_{ij}(x,y) = f_{i}(x) g_{j}(y).
\end{eqnarray}

Neurons in the final output layer perform a weighted sum of sensorimotor inputs and encode the direction that the dragonfly should turn to intercept its prey. The response $R$ of output neuron $k$ is:
\begin{eqnarray}
    R_{k}(x,y) = \beta\sum_{i,j} W_{ijk}S_{ij}(x,y),
\end{eqnarray}
where 
\begin{eqnarray}
    W_{ijk} = \iint \exp\left( -\frac{(y+z-a_{i})^2}{2\sigma_{r}^2}\right)\exp \left( \frac{(y-b_{j})^2}{2\sigma_{g}^2} \right)\exp \left( \frac{(z-c_{j})^2}{2\sigma_{m}^2} \right).
\end{eqnarray}
Here, $c_{k}$ is the turn-direction that most strongly drives neuron $k$ activity,  $\sigma_{m}$ is a parameter that affects the selectivity of output neurons to turn direction, and $\beta$ may be thought of as a synaptic scaling parameter that determines the overall activity level of the output neurons.

The dragonfly-turn direction, $d$, is decoded from output neuron activations:
\begin{eqnarray}
    d = \frac{\sum_{k} c_{k} R_{k}}{\sum_{k} R_{k}}.
\end{eqnarray}




\end{document}